# PTFE treatment by remote atmospheric Ar/O$_2$ plasmas: a simple reaction scheme model proposal

E.A.D. Carbone[1], M.W.G.M. Verhoeven[2], W. Keuning[3], J.J.A.M. van der Mullen[1]

[1] Elementary Processes in Gas discharges (EPG), Department of Applied Physics, Eindhoven University of Technology (TUe), Eindhovenm The Netherlands
[2] Physical Chemistry of Surfaces, Department of Chemical Engineering and Chemistry, Eindhoven University of Technology (TUe), Eindhoven (NL)
[3] Plasma & Materials Processing Group, Dept. of Applied Physics, Eindhoven University of Technology (TUe), Eindhoven (NL)

E-mail: e.a.d.carbone@tue.nl

**Abstract.** Polytetrafluoroethylene (PTFE) samples were treated by a remote atmospheric pressure microwave plasma torch and analyzed by water contact angle (WCA) and X-ray photoelectron spectroscopy (XPS). In the case of pure argon plasma a decrease of WCA is observed meanwhile an increase of hydrophobicity was observed when some oxygen was added to the discharge. The WCA results are correlated to XPS of reference samples and the change of WCA are attributed to changes in roughness of the samples. A simple kinetics scheme for the chemistry on the PTFE surface is proposed to explain the results.

## 1. Introduction

Cold atmospheric pressure plasmas (CAPPs) receive nowadays an ever growing interest [1] in many different applications fields like surface treatments, thin film coatings, plasma medicine, waste residues conversion and ozone production.
In order to improve these applications, insight in the physics and chemistry of these plasmas is needed. However, these CAPPs present additional complexity compared to e.g. classical low pressure plasmas. The creation of radicals and active species is no longer ruled by the electrons solely. Instead, many other excitation paths are important that are realized by interactions between excited heavy particles. Moreover molecular assisted recombination processes may also be important in the creation of radicals. In contrast to (atomic) low pressure plasmas the plasma recombination processes do not take place at the walls but mainly in the volume.
This study is devoted to the treatment of polymers using a CAPP in a remote set-up. Since the plasmas are low in gas temperature and far from maxwellian equilibrium the influence of physical processes like thermal heating and ion sputtering can be neglected. These destructive aspects of plasma treatment are further reduced by operating the CAPP in a remote set-up. Remote cool plasma treatments show numerous advantages as they modify [2] only the top layers of the substrate and do not alter the bulk properties of the material.

To probe the kinetics and to understand the action of CAPPs it is important to choose a model-surface with limited reaction paths so that non-ambiguous surface responses can be traced and studied. As a model-surface we selected PTFE (polytetrafluoroethylene), the basic material of teflon. This is a high thermal resistive polymer and has a monomer unit composed only of C and F. Its chemical structure can be defined as $-(CF_2-CF_2)_n-$. This surface is rather non-reactive and exhibits the property of being hydrophobic. Moreover C and F are both easy detectable by means of X-ray photoelectron spectroscopy (XPS). This surface analysis technique can give the stochiometric composition of the 8-9 first monolayers of the polymer substrate and is able to detect all elements of the periodic table apart H and He. With respect to these considerations, PTFE is a natural choice.

The PTFE samples were treated with an atmospheric pressure microwave surfatron and analyzed afterwards by XPS and water contact angle (WCA) measurements. The results are compared with the available literature. A simple reaction scheme model for the surface reactions is eventually proposed to explain the observed behavior.

## 2. Experimental setup

The plasma is generated by the absorption of electromagnetic waves of 2.45 GHz in an argon flow. The surfatron is a Sairem® launcher coupled to a Microtron™ microwave generator that can deliver powers in the range of 10 to 200W. The Argon and $O_2$ flow-rates inside the ceramic tube were controlled using Brooks mass flow controllers. An air flow of 20L/min was used as extra cooling of the walls of the plasma tube and the plasma itself. The tube has an inner radius of 0.4 mm and an outer diameter of 1.2 mm. The PTFE samples were mounted on a XY linear motorized translator table in the normal direction to the plasma.

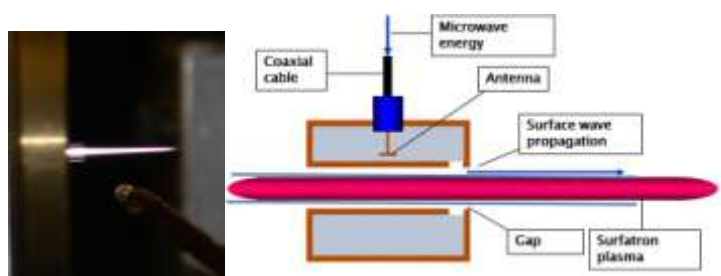

Figure 1 : Picture and schematic of the microwave induced plasma (MIP) torch

The PTFE sheets were purchased from Goodfellow (1 mm thick sheet, Ref FP303050). After cutting the sheets to sizes of 1 cm$^2$, these were cleaned using pure ethanol (Normapur VWR) followed by pure isooctane (GR for analysis, Merck). Subsequently the PTFE samples were treated by the plasma torch. The XPS measurements were carried out with a Kratos AXIS Ultra spectrometer equipped with a monochromatic X-ray source and a delay-line detector (DLD). Spectra were obtained using the aluminium anode (Al Kα) operating at 150W. Survey scans were measured at a constant pass-energy of 160 eV and region-scans at 40 eV. The background pressure was 2 x 10$^{-9}$ mbar.

The WCA measurements were taken in a static mode with a KSV (CAM200) equipment while using distilled 1μl water droplets. The liquid-solid contact angle is a measure of the surface energy of the substrate. For a perfectly flat (at the atomic scale) surface, it can be directly correlated to the composition in chemical groups of the top layers of the sample. However, it also depends on the roughness of the sample. WCA results are then always the convolution of the chemical nature of the surface but also of its morphology.

## 3. PTFE surface treatment results

The PTFE samples were treated by pure Ar and Ar/O$_2$ mixtures and their changes in surface properties were probed by WCA. In table 1, it is shown that in the case of pure argon treatment, the hydrophobic character remains nearly constant for short treatment time and that it *decreases* slightly for longer treatments. The lowest WCA-value that was found is 90°. This is a typical value for PTFE after treatment by a pure Argon plasma and is usually reported as "cleaning the surface" [10]. For untreated PTFE the WCA values are typically ranged in between 90-110° because they depends strongly of the processing method to manufacture the polymer sheets. It shall be noted that the error on the WCA in these measurements was not more than ±2°. Moreover in [3], it was reported that an Argon treatment leads to a slight smoothening of the surface.

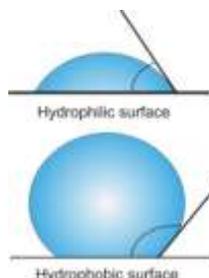

| Power | Distance | Ar/O$_2$ flux [L -mL/min] | Time | WCA |
|---|---|---|---|---|
| Untreated PTFE sample | | | | 110° |
| 25 W | 2 mm | 1,5 – 0 | 600 s | 90° |
| 50 W | 2 mm | 1,5 – 0 | 15 s | 108° |
| 50 W | 6 mm | 1,5 – 35 | 120 s | 117° |
| 50 W | 4 mm | 1,5 – 35 | 120 s | 126° |

*Table 1: Water contact angle of PTFE samples treated by the microwave plasma torch: a decrease of the WCA after Argon plasma treatment is observed. In the case of an Ar/O$_2$ plasma treatment (the two bottoms lines) we an increase in WCA and thus the hydrophobicity.*

When some little amount of Oxygen is added (~0.1 %) in the argon discharge, an *increase* of hydrophobicity of the surface is observed contrary to the pure argon case. A higher WCA-value corresponds to a higher degree of hydrophobicity (lower surface energy). Moreover closer distances lead to higher WCA values. This is in contrast to what is usually expected for oxygen plasma treatments where oxygen grafting leads to chemical groups like C-O, C=O or COO on the surface [4]. The polar groups would lead to a reduction the WCA and this is not what we observe.

To get a better understanding of the surface changes, we performed XPS measurements on the plasma-treated samples. The resulting stochiometric findings of the sample surfaces as function of plasma treatment are given in Table 2.

| Power /Ar-O$_2$ flux/Distance | | | % C | % F | % O |
|---|---|---|---|---|---|
| Untreated PTFE sample | | | 32,1 | 67,9 | 0 |
| 50W | 3, 3L -35mL/min | 3 mm | 31,22 | 67,94 | 0,84 |
| 50W | 3, 3L -35mL/min | 6 mm | 31,46 | 67,74 | 0,8 |
| 50W | 3, 3L -17mL/min | 6 mm | 31,1 | 68 | 0,9 |
| 50W | 3,3L - 0 mL/min | 6 mm | 32,19 | 66,91 | 0,9 |

*Table 2: Atomic composition of PTFE surfaces after treatment by the MIP. The stoechiometry of the samples is conserved both in the case of Ar and Ar/O$_2$ plasmas.*

The table shows that both the Argon and Argon/Oxygen cases do not lead to a change in the C:F ratio. Moreover there is nearly no grafting of oxygen: less than 1%. This is not significant as it is in the error margins of XPS (~1 %).

The deconvolution of the C1s spectra before and after treatment showed no obvious changes in the shape and FWHM of the C1s peak at 292eV. The measured FWHM C1s of PTFE was about 1.2eV before and after the plasma treatments. This indicates that the chemical nature of PTFE − $(CF_2-CF_2)_n-$ is conserved in addition to its stochiometry.

These XPS results lead to a straightforward interpretation of the WCA findings. Since there are almost no chemical changes of the surface the decrease of the hydrophobicity of the surface must be attributed to the decrease of roughness in the case of the pure Argon plasma treatment. Conversely, the increase of hydrophobicity in the case of $Ar/O_2$ plasma can be associated to an increase of roughness of the samples [3, 5].

## 4. Proposal of a chemical model for the surface kinetics

It is instructive to compare the results of the current study done with a atmospheric *microwave* induced plasma with those obtained by a *RF* atmospheric pressure plasma. In [3], Carbone et al. showed that the addition of oxygen to the RF discharge does not lead to an *increase* but a *decrease* of the surface energy of the PTFE surface. Also in that study, no significant O grafting on the surface was observed by XPS. Moreover the latter indicated towards a conservation of the chemical nature of the polymer. Last but not least, it was found that the roughness of the samples was also increased in the case of $Ar/O_2$ and this roughening/etching of the surface was confirmed by atomic force microscopy. This is exactly the same kind of behavior that we report here but now for a different plasma source and configuration. Similar trends were also shown in [8,9].

To confirm that the thermal loading of the substrate by means of the plasma cannot explain the changes in trends, we performed Rayleigh scattering on this microwave plasma [19]. The temperature right at the outlet, is found to be as high as 800 K but further downstream decreases down to about 400K. When $O_2$ is added to the discharge, and using the same technique, it is found that the gas temperature remains constant or even decreases [6] from 750 to 700 K right at the exit of the tube for flow rates from 0 up to 15 sccm in 1 L/min of Ar. The polymers were treated downstream in the region where the plasma was cold (~400K).

The similarity of the results of these two studies, the current one and the one reported in [3] leads to the idea that we face some kind of 'universal' behavior that can be explained by the 'remote' and 'radical chemistry driven' nature of the $Ar/O_2$ afterglow in contact with surfaces.

To understand this 'universal' behavior a simple reaction scheme model was constructed based on a division of the process in three steps 1) activation, 2) re-association and 3) chemical etching.

*4.1. Pure Argon chemistry*

The first step, the activation step, leads, in any plasma-surface interactions, to the breaking of some bounds on the surface. This provides the creation of radicals and allows them to react with each other and/or with radicals coming from the plasma. For the PTFE plasma treatment we can define two main roads: one involving the breaking of C-F bounds and the other the breaking of C-C bounds:

- an example of C-F bound breaking is the production of negative fluorine ions by electron dissociative attachment [7]:

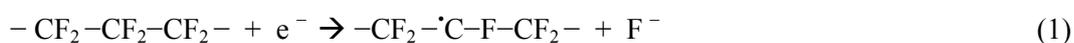

$$- CF_2-CF_2-CF_2- + e^- \rightarrow -CF_2-{}^{\bullet}C-F-CF_2- + F^- \qquad (1)$$

However in the conditions of this study, defluorination seems negligible as no change in the C:F stochiometry is observed with is its usual signature in fluorinated polymers plasma treatments.

- an example of the C-C breaking is the *carbon backbone breaking* by homolytic (formation of two radicals due to one bound breaking) dissociation. This can be induced by photons, electrons and/or Ar* metastables [8]:

$$-CF_2-CF_2- + h\nu \text{ (or e-, Ar*)} \rightarrow -CF_2^\bullet + {}^\bullet CF_2- + \text{(e-, Ar)} \qquad (2)$$

This step will be of main importance for the surface activation and etching as defluorination is negligible, carbon atoms become accessible for surface reactions through C-C bounds breaking.

In the pure Argon plasma case, the recombination of fluorine radicals on the surface can occur [9-10]:

$$-CF_2- ({}^\bullet C-F) -CF_2- + F^\bullet \rightarrow -CF_2- (F-C-F) -CF_2- \qquad (3)$$

$$-CF_2-CF_2^\bullet + F^\bullet \rightarrow -CF_2-CF_3 \qquad (4)$$

And more importantly, we will have recombination of fluorocarbon radicals between them directly on the surface [11]:

$$-CF_2-{}^\bullet CF_x + {}^\bullet CF_y \rightarrow CF_2-CF_x-CF_y- \qquad (5)$$

Although no clear evidence came from our measurements that these processes take place, it cannot be ruled out, moreover it was observed by different authors [3, 9, 14].

*4.2. Argon/Oxygen chemistry*

The activation steps will be comparable to the pure argon case, meaning that the activation is mainly driven by electrons, photons and Argon metastables rather than by oxygen atoms/molecules. The reason for this assumption is that the ground state $O_2$ density is only 0.1% of that of argon. However, in the case of this atmospheric pressure plasma jet the dissociation degree of $O_2$ is unknown but might be expected to be quite high [18]. Nevertheless the emission spectrum of the plasma is still largely characteristic of a pure Argon plasma so that we can expect that Argon metastables are still one of the dominant species in the plasma. In the case of $Ar/O_2$, the main chemistry that occurs is between the radicals formed on the surface with the species coming from the gas phase. The fluorocarbon radicals can either react with atomic oxygen [11] to form *alkoxydes*:

$$-CF_2- ({}^\bullet C-F) -CF_2- + O \rightarrow -CF_2- (F-C-O^\bullet) -CF_2- \qquad (6)$$

or carbonyles [11]:

$$-CF_2-C^\bullet-CF_2- + O \rightarrow -CF_2-(C=O)-CF_2- \qquad (7)$$

Another reaction path is the direct reaction or the fluorocarbon radicals with molecular oxygen to form *peroxydes* [12-13]:

$$-CF_2-CF_2^\bullet + O_2 \rightarrow -CF_2-CF_2O_2^\bullet \qquad (8)$$

which can also recombine to form *alkoxydes* [13]

$$2 (-CF_2-CF_2O_2^\bullet) \rightarrow 2 (-CF_2-CF_2O^\bullet) + O_2 \qquad (9)$$

This second path is believed to be of higher importance compared to the steps (6) and (7) because of the large reservoir of $O_2$ available for reacting on the surface. This is even more likely far in the plasma afterglow where O atoms are expected to recombine to form $O_2$ and $O_3$ in the gas phase.

Finally, the intermediates of reaction produced at the steps (6), (7), (8) and (9) can lead to the production of different radicals

$$-CF_2 - CF_2 - CF_xO_y^{\bullet} \rightarrow -CF_2 - CF_2^{\bullet} + (CO + CO_2 + CF_2O + CF_2) \quad (10)$$

and other gaseous species as shown in [11-12,15-16] which leave the surface. In that way, they do not contribute anymore to the grafting of C-O$^{\bullet}$, C=O or COO$^{\bullet}$ species on the surface. We have then conservation of the *PTFE-like* character of the polymer after treatment. According to our results, this last step needs to be fast compared to the activation and (partial) oxidation of the carbon atoms. Indeed, this is the only way how, one can expect conservation of the stoechiometric composition of the surface. The chemical etching of the surface will lead ultimately to an increase of roughness of the Teflon sample while oxygen is added to the argon plasma.

## 5. Conclusions

In this study, we showed that PTFE surfaces can be modified in terms of roughness meanwhile keeping the surface chemistry and stoichiometry nearly unchanged. This behavior seems characteristic of remote $Ar/O_2$ plasma treatment at atmospheric pressure. A simple three phase chemical scheme is proposed for the interactions between the plasma and the PTFE surface. The first step consists of an activation of the surface by the argon plasma. This is followed by a second step of soft chemical etching of the carbon backbone through radical reactions mainly with molecular oxygen. The fast desorption of the newly formed $C_xF_yO_z$ radicals in the third step allows then finally to obtain no significant changes in the surface stochiometry. Species with a x:y ratio of 0.5 seems to be the ones which leave preferentially the surface with variable amount of O.

Although a complete and exact description of the plasma/surface kinetics would require a full 2D modeling and to implement a complete chemistry model for the $Ar/O_2$ plasma and its interactions with $CF_2$-$CF_2$, a simple approach based on conservation laws was preferred here. The authors would like to point out here the uphill task to find accurate rate coefficients for the surface reactions. They depends strongly of diffusion of chains which depends itself of the length of the polymer chains which is not a conserved parameter during the etching. This simple approach could be extended for other reactive mixtures like $Ar/N_2$ and $Ar/CO_2$ which are of main importance in surface treatment and especially for biomedicine applications [17].